\documentclass{appolb}
\usepackage{epsfig}
\newcommand{\nc}{\newcommand}
\nc{\ds}{\displaystyle}        \nc{\ts}{\textstyle}
\nc{\rf}[1]{Fig.\,\ref{#1}}    \nc{\rt}[1]{table\,\ref{#1}}
\nc{\req}[1]{Eq.\,(\ref{#1})}  \nc{\eps}{\varepsilon}
\nc{\beq}{\begin{equation}}     \nc{\beql}[1]{\begin{equation}\label{#1}}
\nc{\eeq}{\end{equation}}
\nc{\beqa}{\begin{eqnarray}}   \nc{\eeqa}{\end{eqnarray}}
\nc{\bfi}{\begin{figure}}       \nc{\efi}{\end{figure}}
\begin{document}
\title{RESONANCES DO NOT EQUILIBRATE}
\author{Inga Kuznetsova$^1$, Jean Letessier$^{1,2}$, and Jan Rafelski$^{1,3}$
\address{$^1$Department of Physics, University of Arizona, Tucson, AZ 85721}
\\and
\address{$^2$Laboratoire de Physique Th\'eorique et Hautes Energies\\
LPTHE, Universit\'e Paris 7, 2 place Jussieu, F--75251 Cedex 05}
\\and
\address{$^3$Department f\"ur Physik der Ludwig-Maximillians-Universit\"at M\"unchen und\\
  Maier-Leibniz-Laboratory, Am Coulombwall 1, 85748 Garching, Germany}
}
\maketitle
\begin{abstract}
We discuss, in qualitative and quantitative fashion, the
 yields of hadron resonances. We show that these
yields, in general, are not in  chemical equilibrium.
We evaluate the non-equilibrium abundances 
in a dynamic model  implementing  the  $1+2\leftrightarrow 3$ 
resonance formation reactions. Due to the strength of these reactions, we show
the $\Sigma(1385)$ enhancement, and the $\Lambda(1520)$ 
suppression  explicitly.
\end{abstract}

\section{Why study strange resonances and quark--gluon plasma?}
We study the  quark  confining  vacuum structure, with the  objective
to   liberate quarks and gluons at high temperature.
The present day experimental effort involves  colliding heavy nuclei and this   limits
the domain of deconfinement to the   size of the atomic nucleus.
 The `free' quarks and gluons form a thermal gas comprising color
charges, the quark--gluon plasma ---  a hot soup of elementary  matter,
last seen in the Universe  when matter was formed at about
30 $\mu$s after the big-bang.

The temperature at which we deconfine the quark content of nucleons is  $T_c\simeq 160$ MeV. However,
we reach in heavy ion collisions, $T_{\rm Max}\simeq 2$--$5\,T_c$. At this relatively high temperature,
we create strange quark and antiquark pairs in an abundance which rivals
that of light quarks~\cite{Rafelski:1980rk}. While this is going on, the compressed quark matter
expands, the expansion consumes the thermal pressure and energy,
 and ultimately the  fireball of hot quarks breaks apart at $T_f<T_c$.  Of particular
interest is the  high abundance  of antiquarks, including anti-strangeness.
The high strange antimatter yield is our evidence that we have recreated
the early Universe in  a laboratory set experiment.

We are interested in understanding the physics of quark--gluon plasma in the last moments
of its existence, and thus the conditions prevailing in the transformation of
quarks and gluons into hadrons.  The yields of  particles produced can be successfully
described using statistical physics methods. We use the program SHARE
(Statistical HAdronization with REsonances) for this purpose~\cite{Torrieri:2004zz,Torrieri:2006xi}.
The remarkable  success of this model  indicates that the production of hadrons in heavy ion
collisions is governed mostly by the accessible phase space. This, in turn, implies
that particles are produced in a process resembling vapor evaporation from a hot soup. In this 
process, hadrons are formed from  very `sticky' quarks and this helps to saturate  the probability of
particle formation.

Aside of stable (under strong interactions) particles, the statistical
hadro\-nization  of quark--gluon plasma predicts the
yields  of  (anti)baryon resonances.
Experimental results available today show that the yields of these states do not
always follow  the  model
expectations~\cite{Broniowski:2003ax,Markert:2002xi,Adams:2006yu,Salur:2006jq,Markert:2007qg,Witt:2007xa,Abelev:2008yz}.
Given the success of SHARE, we interpret
this as  a post hadronization evolution of observable yields~\cite{Rafelski:2001hp,Bleicher:2002dm}. Thus,
the first objective of this report, see section~\ref{SecEqu}, is to explain
why, for stable particles, we can directly apply the statistical
hadronization model, while observed resonance yields are subject
to  post-hadronization dynamics.

After that,
we develop the kinetic model tools in section~\ref{secEvol}, define the model and 
the initial conditions in section~\ref{SecModel}, and present, in the following
section~\ref{SigLam},
a detailed numerical study demonstrating that  $\Sigma(1385)$ can be greatly enhanced
 in the observed abundance  compared to statistical  equilibrium,
while other more stable resonances, such as $\Lambda(1520)$ are suppressed.

The understanding of this behavior of
strange baryons and antibaryons and their resonances
 sharpens the tools available to us  in the study of
quark--gluon plasma properties, at the time of phase conversion into hadrons.
This work improves the  understanding of the physics
of hadronization of quark--gluon plasma, that is of the process of freezing of the
deconfined vacuum.

\section{Why resonances, in general, do not chemically equilibrate?}\label{SecEqu}
The observed stable particle yield is controlled solely by freeze-out temperature,
and this yield contains the decay of all resonances. However,
the resonance abundances can evolve and mix with stable particles
without altering the observed final stable particle yield, since there is no
information in the stable particle yield about `sharing' of the yield with resonances.
Therefore, we cannot assume  that resonance yields are governed by same physics as
the final stable particle yields. We
consider, as an example, the
reaction
\begin{equation}
\label{piND}
 \Lambda + \pi  \leftrightarrow \Sigma^*,
\end{equation}
and imagine, for purpose of following simple illustration, that the system we study comprises ONLY
these three particles.

Reaction Eq.\,(\ref{piND}) does not change either the pion $\pi$ or lambda  $\Lambda$ yield,
since all $\Sigma^*$-resonances ever made will
ultimately contribute to these yields upon resonance decay, which naturally happens before
the stable particles are observed. On the other hand,  we  measure the yield of $\Sigma^*$ by the
invariant mass method. This is done by considering all pair combination of two presumed
decay products and evaluating from the energy and momentum  of both particles the  invariant
mass distribution $dN/dM$ where:
$$M=\sqrt{(E_\pi+E_\Lambda)^2-(\vec p_\pi+\vec p_\Lambda)^2}.$$
This method implies that
as long as the decay products do not rescatter after decay, that is
before leaving   the medium, the yield
of resonances is determined by the observed yield of the decay products.
On the other hand, we can assume that each elastic scattering deflects the
momentum vector, so that the invariant mass method fails to observe
the resonance.

Given this consideration, we can evaluate the relative yield of $  \Sigma^*/ \Lambda$
assuming that the system expands very slowly (and we ignore   spin and isospin  for simplicity). 
We are given  the inelastic reaction rate, $R(T)_{\rm in}$, at temperature
$T$ derived from the
cross section governing the inelastic  reaction Eq.\,(\ref{piND}).  Similarly, we  have
total elastic rate, $R(T)_{\rm el}$,   originating in any
elastic   scattering (that is of $\pi$ or $\Lambda$)  in the medium.
The elastic reaction rate is proportional
to the probability of not observing a $\Sigma^*$, while
the inelastic reaction rate, with the
same proportionality constant, is describing the probability of
seeing a $\Sigma^*$.  This means that   the {\em invariant mass method
observed} relative $\Sigma^*$ yield is the ratio of inelastic scattering rate to any
reaction rate in the medium, that is,
\begin{equation}\label{ratio}
\frac{\Sigma^*}{\Lambda}=\frac{R(T_r)_{\rm in}}{R(T_r)_{\rm in} +R(T_r)_{\rm el}},
\end{equation}
considering that the nucleon yield already comprises all $\Sigma^*\to \Lambda$ decays.
Here, we denote by $T_r$ the resonance free-streaming  temperature which is typically
lower than the stable particle freeze-out condition $T_f$. Note that if the 
elastic reactions are negligible, in this greatly simplified model
all $\Lambda$ are descendants of $\Sigma^*$
and the ratio is unity.

\begin{figure}[tb]
\centerline{\hspace*{0.5cm}\psfig{width=13cm,figure=  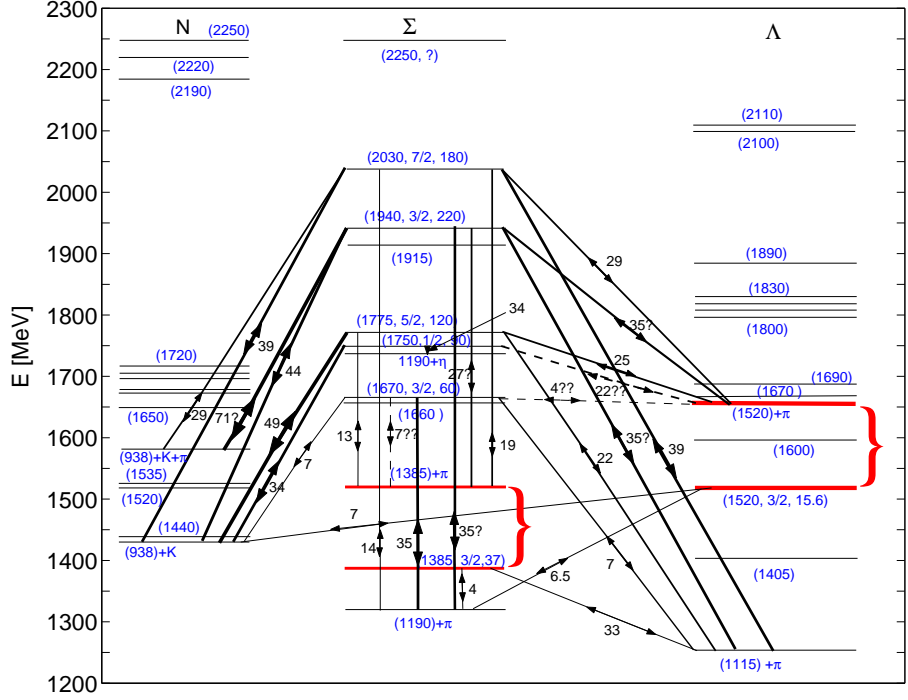}}
\caption{\label{Lam1520}
Reactions scheme for $\Lambda(1520)$ and $\Sigma(1385)$ population evolution.}
\end{figure}

We see, in this example, that the observed  relative resonance yield
is not governed by thermal/statistical
properties of the medium,   but depends decisively on the
strength of the inelastic reaction rate derived from known
resonance cross section of reactions, such as  Eq.\,(\ref{piND}).
In this work, we will be addressing realistic
physical system and will focus our interest on the understanding
of the role of  the inelastic reaction rate of important  resonances.
We will be following their abundance evolution as function of time $t$, that is for
a given function,   $T(t)$, of temperature.  We thus will present our results
as a function of $T$ and focus interest on the range  $T_r$ to $T_f$,
where $T_f$ is the hadron formation temperature in QGP breakup.

\section{Time evolution equations }\label{secEvol}

In figure~{\ref{Lam1520}}, we show the scheme of reactions  which all
have a noticeable effect on $\Lambda(1520)$ yield after the chemical
freeze-out kinetic phase. The format of this presentation is
inspired by nuclear reactions schemes. On the vertical axis, the
energy scale is shown in MeV. There are three classes of particle
states, which we denote from left to right as `$N$' (S=0 baryon),
`$\Sigma$' ($S=-1,\ I=1$ hyperon) and `$\Lambda$' ($S=-1,\ I=0$
hyperon).

 Near each particle  bar,   we state (on-line in blue) its mass,
and/or angular momentum, and/or total width in MeV. The states
$\Lambda(1520)$ and $\Sigma(1385)$ are shown along with the location
in energy of $\Lambda(1520)+\pi$ and $\Sigma(1385)+\pi$
respectively, both entries are connected by the curly bracket, and
are highlighted (on-line in red). The  inclusion of the $\pi$-mass
is helping to see the kinetic threshold energy of a reaction. The
lines connecting the $N,\ \Sigma$ and $\Lambda$ columns are indicating the
reactions we consider in the  numerical computations. All reactions
shown in figure~{\ref{Lam1520}} can go in both directions, as shown
by the double arrows placed next to the numerical value of the
partial decay width $\Gamma_i$, in MeV.

The  evolution, in time,  of the resonance yield is described
by the process of resonance formation in scattering, $ 1 + 2 \rightarrow 3$,
less natural decay $3\rightarrow 1 + 2  $:
\begin{equation} \label{delev}
 \frac{1}{V}\frac{dN_{3}}{dt}=\sum_i\frac{dW^i_{{1+2 \rightarrow
3}}}{dVdt}-\sum_j\frac{dW^j_{{3 \rightarrow 1 +2}}}{dVdt},
\end{equation}
where subscripts $i$, $j$ denote different reactions channels when
available. We
further allow different sets of subscripts $i$, $j$ in order to allow
more complex dynamical  cases in which not all production and/or decay
channels are present.

Allowing for Fermi-blocking and Bose enhancement in the final state,
where  by designation   particles $1$  and $3$ are fermions (heavy baryons) and
particle $2$ is a boson (often light pion), we have for the two rates:
\begin{eqnarray}
\frac{dW^j_{3 \rightarrow 1 +2 }}{dVdt}=
&&\hspace*{-.4cm}  \int\!\!\frac{g_{3}d^{3}p_{3}}{2E_{3}(2\pi)^3}
     \int\!\!\frac{d^{3}p_{1}}{2E_1(2\pi)^3}\int\!\!\frac{d^{3}p_{2}}{2E_{2}\left(2\pi\right)^{3}}
\left(2\pi\right)^{4}
\delta_p^{4}\left({1}+{2}-{3}\right) \nonumber\\[0.3cm]
&&\hspace*{-.6cm}
\times f_{3}\left(1 - f_{1}\right)\left(1 + f_{2}\right)
\frac{1}{g_{3}}\sum_{\rm spin}\left|\langle p_{3}\left|
M^j\right|p_{1}p_{2}\rangle\right|^{2}\label{dr},
\end{eqnarray}
and in analogy, we have for the $3$ back-production rate
\begin{eqnarray}
\frac{dW^i_{1+2 \rightarrow 3}}{dVdt}&=&
\int\!\!\frac{g_1d^{3}p_{1}}{2E_{1}(2\pi)^3}
\int\!\!\frac{g_{2}d^{3}p_{2}}{ 2E_{2}(2\pi)^3}
\int\!\!\frac{d^{3}p_{3}}{2E_{3}\left(2\pi\right)^{3}}
\left(2\pi\right)^{4}\delta_p^{4}\left({1}+{2}-{3}\right)
  \nonumber\\[0.3cm]
&&\hspace*{-.4cm}
\times f_{1}f_{2} \left(1 - f_{3}\right)
 \frac{1}{g_1g_{2}}\sum_{\rm{spin}}\left|
\langle p_{1}p_{2}\left| M^i\right|p_{3}\rangle\right|^{2},
\label{pr}
\end{eqnarray}
where 
$\delta_p^4(1+2-3)\equiv\delta^4(p_1+p_2-p_3)$ assures
4-momentum conservation   and $g_i,\ i=1,\ 2,\ 3$ is 
particle  degeneracy. The Bose 
function  for particle  $2$, and Fermi  distribution for particles $1,\ 3$ are:
\begin{eqnarray}
f_{2} = \frac{1}{\Upsilon_{2}^{-1}e^{u\cdot p_{2}/T} - 1},\qquad 
%
f_{j} = \frac{1}{\Upsilon_j^{-1}e^{u\cdot p_j/T} + 1},\ j=1,\ 3\label{fN}.
\end{eqnarray}
Here, $\Upsilon_i$ is particles fugacity, and  $u\cdot p_i=E_i$, for
$u^\mu=(1,\vec 0)$ in the rest frame of the heat bath where
 $d^4p\delta_0(p_i^2-m_i^2)\to d^3p_{i}/E_{i}$ for each particle. Hence,
Eq.(\ref{dr}) and Eq.(\ref{pr}) are  Lorentz invariant, and thus as presented these rates
can be evaluated  in any convenient frame of reference. Normally, this is the
frame co-moving with  the thermal volume element.

Since   particles  2, 3 are here   heavy baryon  (resonances),
 we can work using the expansion of the
relativistic  distribution, the first term is the Boltzmann limit:
\begin{eqnarray}
\frac {N_{i} }V = \Upsilon_{i}\frac{T^3}{2\pi^2}g_{i}x_{i}^2K_2(x_{i}), \label{relboltz}
\end{eqnarray}
where $x_{i}=m_{i}/T$, $K_2(x)$ is Bessel function (not to be mixed
up with particle~2). However, we use the complete Bose distribution
to describe pions.

We introduce in medium lifespan of particle 3:
\begin{equation}
\frac{1}{\tau_3}\equiv   \frac{\sum_i R^i_{123}}{V^{-1}dN_3/d\Upsilon_3},\label{Dect}
\end{equation}
and, similarly, channel lifespan $\tau_3^i$, omitting the sum $\sum_i$.
The rate $R_{123}$ is:
\begin{eqnarray}
R_{123}^i=
\int\!\! \!\!\int\!\! \!\!  \int\!\!
 \frac{d^3p_1d^3p_2d^3p_3}
         {8E_1E_2E_3(2\pi)^5}
\frac{f_1 f_2   f_3\,e^{u\cdot p_3/T}}{\Upsilon_1 \Upsilon_2 \Upsilon_3 }
    \delta_p^{4}\left({1}+{2}-{3}\right)
\sum_{\rm{spin}}\left| \langle {1}{2}\left| M^i\right|{3}\rangle\right|^{2}.
\label{prR}
\end{eqnarray}
$R$  is independent of the fugacity, in the Boltzmann-limit.

The production and decay rates are connected to each other by the
detailed balance relation~\cite{KuznKodRafl:2008,Kuznetsova:2008jt}:
\begin{equation}
\Upsilon_1^{-1} \Upsilon_2^{-1}\frac{dW_{1+2 \rightarrow 3}}{dVdt}=
 \Upsilon_{3}^{-1} \frac{dW_{3 \rightarrow 1+2}}{dVdt}=R_{123}.\label{pdr}
\end{equation}
Using detailed balance Eq.\,(\ref{pdr}), we obtain  for fugacity
$\Upsilon_3$ the evolution
equation:
\begin{equation}
\frac{d\Upsilon_{3}}{d{\tau}} =
\sum_i{\Upsilon^i_{1}\Upsilon^i_{2}}\frac{1}{\tau^i_{3}}
+\Upsilon_{3}\left(\frac{1}{\tau_T}+\frac{1}{\tau_S}-\sum_j\frac{1}{\tau^j_3}\right),
\label{Ups2}
\end{equation}
where we have also introduced  characteristic time constants 
of temperature $T$ and entropy $S$ evolution
\begin{eqnarray}
\frac{1}{\tau_{T} } =  -\frac{d\ln({x_{3}}^2K_2(x_{3}))}{dT} \dot{T}, \qquad 
\frac{1}{\tau_{S} } = -\frac{d\ln( VT^3  )}{dT} \dot{T}. \label{Seq}
\end{eqnarray}
The entropy term is negligible, $\tau_S\gg \tau_3,\ \tau_T$ since we
implement near conservation of entropy during the 
expansion phase. We implement this in way
which would be exact for massless particles taking  $VT^3=$ Const..
Thus, there is some entropy growth in HG evolution we consider, but
it is not significant. In order to evaluate  the magnitude of
$\tau_T$, we use the relation between Bessel functions of order 1 and
2 (not to be mixed up with particles 1, 2)
${d}\left(z^2K_2(z)\right)/{dz}=z^2K_1(z)$. We obtain
\begin{equation}
\frac{1}{\tau_{T}} =  \frac{K_1(x_{3})}{K_2(x_3)}x_3 \frac{\dot{T}}{T}, \label{Teq2}\\
\end{equation}
$\tau_{T}>0$.
For a static system with $\tau_T \to 0$, we see that Eq.\,(\ref{Ups2})
has transient stable population points whenever
\begin{equation}
\sum_i\Upsilon_1^i\Upsilon_2^i/\tau_3^i-\Upsilon_3\sum_j1/\tau_3^j=0. \label{stable}
\end{equation}

Finally, we consider the evolution in time of $\Upsilon_{1}$ and $\Upsilon_{2}$.   In the equation
for $\Upsilon_{1}$, we have terms which compensate what is lost/gained in  $\Upsilon_{3}$
see Eq.\,(\ref{Ups2}). Further, we have to allow that particle `1' itself plays the role
of particle `3' (for example, this is clearly the case for $\Lambda(1520)$).
That is accomplished introducing  a chain of populations relations as follows:
 \begin{equation}
(1'+2' \leftrightarrow 1) + 2 \leftrightarrow  3. \label{dpr2}
 \end{equation}
 In the present
setting,  $\Upsilon_{2=\pi}=$ Const.. By virtue of entropy conservation
  the same applies to the case $2'=\pi$. However,
if either particle $2$ or  $2'$ is a kaon,  we need to follow the equation for
$\Upsilon_{2,2'=K}$ which is analogous  to equation for particle $1$ or $1'$.

\section{Model details and initial conditions}\label{SecModel}
The evolution equations can be integrated once  the time dynamics of
the fireball and the initial conditions are fixed:
\\{\bf 1)} We choose a model of expansion which fixes the behavior $T(t)$; here,
 we  invoke a model of matter expansion where the longitudinal
and transverse expansion is considered to be (nearly) independent:
\begin{equation}\label{DTT}
\frac{\dot {T}}{T} = -\frac{1}{3}\left( \frac{2\,(v\tau/R_{\perp}) + 1}{\tau}\right),
\end{equation}
where $R_{\perp}$ is the transverse radius, $v$ is the velocity of expansion
in transverse dimension. All flow parameters (or temperature
dependence on $\tau$) are the same as in~\cite{Kuznetsova:2008zr,Kuznetsova:2008hb}.
\\{\bf 2)} We determine the {\em initial}
values of particle densities (fugacities) established at
hadronization/chemical freeze-out.
We determine these for RHIC
head-on Au--Au collisions at $\sqrt{s_{\rm NN}}=200$ GeV. We
introduce  the  initial hadron yields  inspired by a picture of a
rapid hadronization of QGP in which quarks combine into final state
hadrons. For simplicity, we assume that  the net baryon yield at
central rapidity  is negligible. Thus, the baryon-chemical and
strangeness potentials vanish. The initial yields  of mesons ($q\bar
q, s\bar q$ and baryons $qqq,\ qqs$ are controlled aside of the
ambient temperature $T$, by the  constituent light quark fugacity
$\gamma_q$ and the strange quark fugacity $\gamma_s$.
\\{\bf 3)} Since high energy pions are moving faster than the bulk of the 
matter and leave the domain in which the slower baryons are found, 
we assume that it is impossible to excite reactions with high threshold 
energy.  We thus exclude channels for resonance 3 production with threshold energy
$\Delta E > 300$ MeV. 
\\{\bf 4)} We characterize the hadronization dynamics: 
we assume that
the strange\-ness pair-yield in QGP is maintained in transition to HG.
This fixes the initial value of $\gamma_s$. In fact, since we
investigate relative chemical equilibrium reactions, our results
do not depend significantly  on the exact initial value $\gamma_s$
 and/or strangeness content. The entropy
conservation at hadronization fixes $\gamma_q$. For hadronization
temperature $T(t=0)\equiv T_0=180$ MeV,  $\gamma_q=1$. However, when
$T_0<180$ MeV, $\gamma_q>1$ in order to have entropy conserved at
chemical freeze-out. At $T_0=140$ MeV, $\gamma_q=1.6$ that is close
to maximum possible value of $\gamma_q$, defined by Bose-Einstein
condensation condition \cite{Kuznetsova:2006bh}.
\\{\bf 5)} The initial particle yields are fixed in terms of fugacities:
\begin{equation}
\Upsilon^0_{(1=Y)}=\gamma_q^{2}\gamma_s,  \qquad 
\Upsilon^0_{(2=\pi)}=\gamma_q^{2},  \label{upinpi}
\end{equation}
or
\begin{equation}
\Upsilon^0_{(1=N)}=\gamma_q^{3}, 
\Upsilon^0_{(2={\rm K)}}=\gamma_q\gamma_s, \qquad \label{upinpi1}
\end{equation}
where $Y\equiv \Sigma,\ \Lambda$
is a hyperon, the particle 1   is a baryon and particle 2
is a meson. The particle 3 is always a strange baryon:
\begin{equation}
\Upsilon^0_{(3=Y)}=\gamma_q^{2}\gamma_s. 
\end{equation}
Note that  for  $\gamma_q > 1$,  we have always initially
\begin{equation}
\left.\frac{\Upsilon_1
\Upsilon_2}{\Upsilon_3}\right\vert_{t=0}=\gamma_q^2 \ge 1\,.
\end{equation}
As a consequence, initially the pair of particles 1, 2 reacts into 3.
\\{\bf 6)} 
We do not need to follow the evolution in time for the
pion yield, which is fixed by conservation of entropy per unit rapidity, as
incorporated in Eq.\,(\ref{DTT}). Thus, it is (approximately) a
constant of motion. This can be seen recalling that the  entropy per
pion is nearly 4 within the domain of temperatures considered.
The conservation of entropy implies that pion number is conserved,
and with $VT^3\simeq \mathrm{Const.}$. This further implies that,
during the expansion,
$$\Upsilon_\pi=\gamma_q^2=\mathrm{Const.},$$
which we keep at the initial value.


\section{$\Sigma(1385)$ and $\Lambda(1520)$ yield results}\label{SigLam}
\begin{figure}[htb]
\centerline{\psfig{width=6cm,height=8.cm,figure=  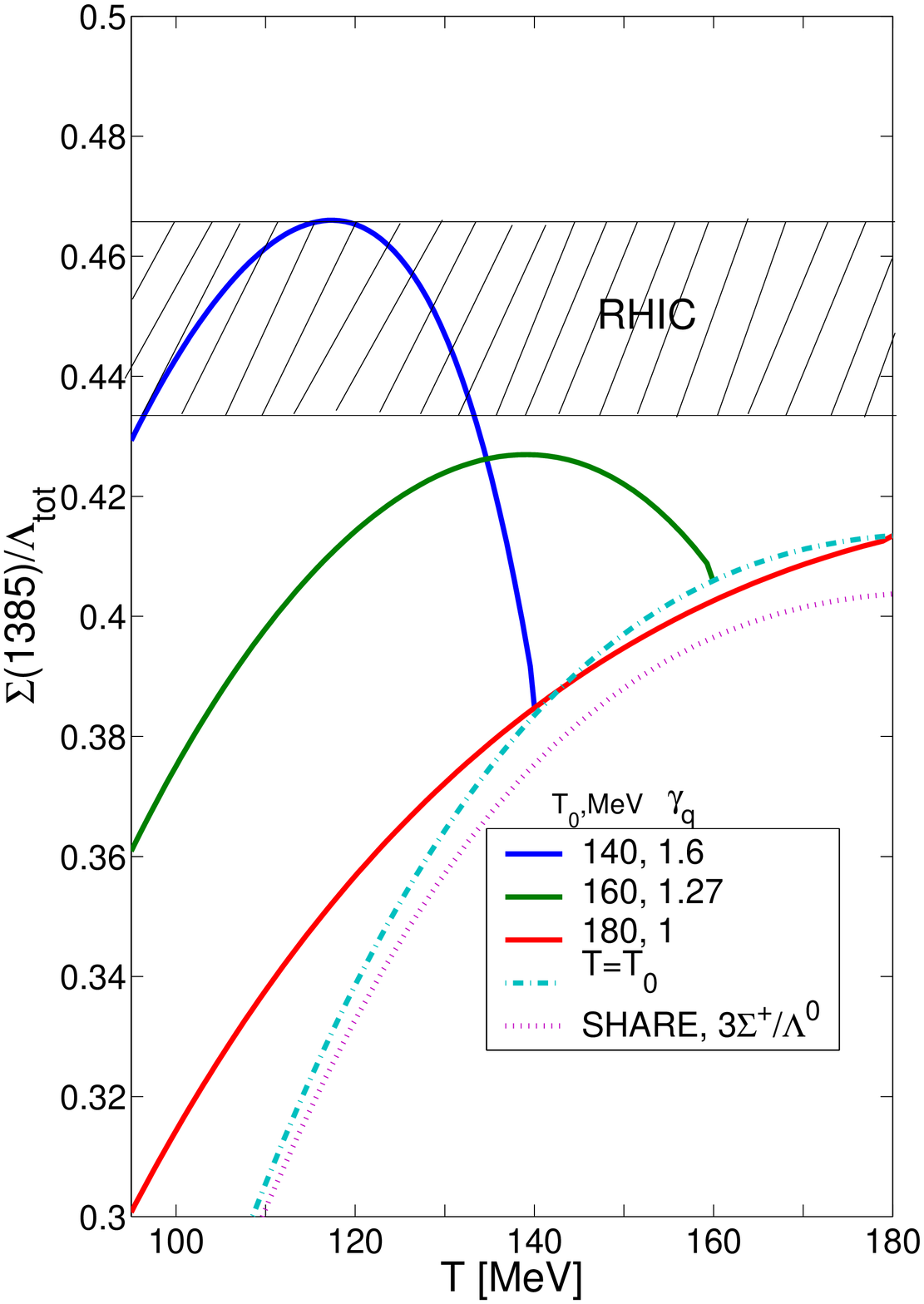}
\hspace*{0.cm}\psfig{width=6cm,height=8.cm,figure=  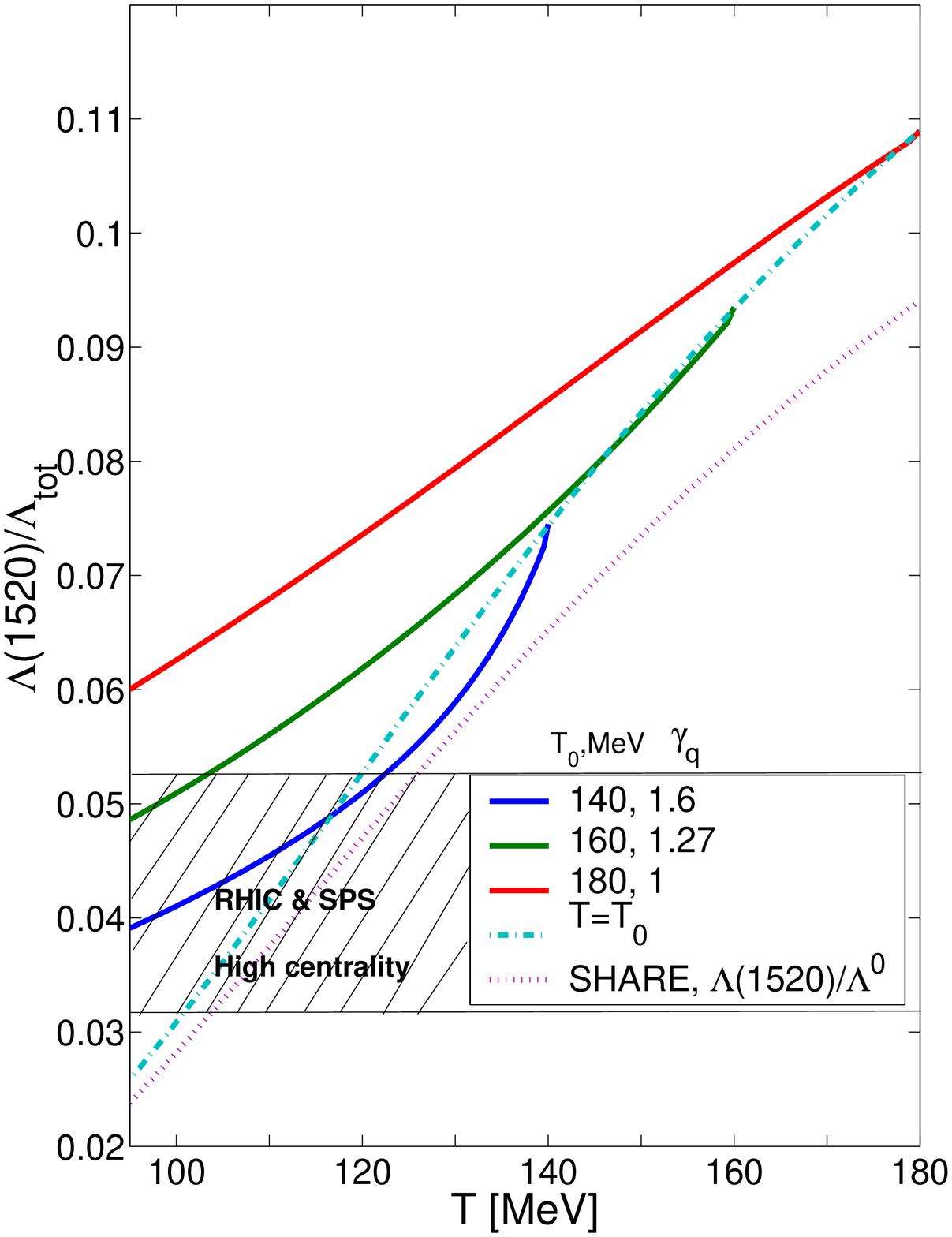 }}
\caption{\label{lamltot}
The ratio $\Sigma(1385)/\Lambda_{\rm tot}$, on left,
and $\Lambda(1520)/\Lambda_{\rm tot}$, on right,
as a functions of temperature $T(t)$ for
different initial hadronization temperatures $T_0=140$, $160$
and $180$ MeV (blue, green and red  lines, respectively, recognized also by 
their initial value along the hadronization curve (dot-dashed).
}
\end{figure}

In  figure~\ref{lamltot}, we present the fractional yields
${\Sigma(1385)}/\Lambda_{\rm tot}$ (left), and
${\Lambda(1520)}/\Lambda_{\rm tot}$ (right) as a function
of temperature of final kinetic freeze-out $T$.
The results  for the hadronization
temperatures $T_0=140$ (blue lines), $T_0=160$ (green
lines) and $T_0=180$ MeV (red lines) are shown.  

In  figure~\ref{lamltot},  the green dash-dotted line is the result
when the kinetic freeze-out temperature $T$ coincides with the
hadronization temperature $T_0$. There is no kinetic resonance
 evolution phase in this
case, only   resonances decay after hadronization. This result is
similar to SHARE result (purple, dotted line). The  small difference
is mainly due to us taking into account the decays
\begin{equation}
\Sigma(1670, 1750) \rightarrow \Lambda(1520) + \pi,
\end{equation}
which are expected/predicted in~\cite{Cameron:1977jr}.
Similarly, for $\Sigma(1385)$ our results for $T_0=T$
are different from SHARE results because we include the decay
\begin{equation}
\Sigma(1670) \rightarrow \Sigma(1385) + \pi,
\end{equation}
which is  part of recently updated particle data set~\cite{Amsler:2008zz}.

For all initial hadronization temperatures, as the freeze-out 
temperature decreases, the suppression for
$\Lambda(1520)_{\rm ob}/\Lambda_{\rm tot}$ ratio is larger than for
$\Lambda(1520)/\Lambda(1520)_0$ (at the same temperature $T$ of final kinetic freeze-out).
The effect is due in part to the asymmetry to excite  $\Sigma(1775)$ and heavier 
 $\Sigma^*$ directly due to its high mass. 
The  resonance yield suppression effect is approximately of the
same magnitude  for all hadronization temperatures $T_0$. However,
the initial hadronization yield of  $\Lambda(1520)$ is sensitive to
temperature, and decreases rapidly with $T$. Therefore, only for $T_0
= 140 $ MeV, a kinetic freeze-out temperatures $\approx$ 95--105
MeV  the  ratio $\Lambda_{\rm ob}(1520)/\Lambda_{\rm tot}$ 
reaches the experimental domain
$\Lambda_{\rm ob}(1520)/\Lambda_{\rm tot}<0.042\pm
0.01$~\cite{Markert:2002xi,Adams:2006yu} shown, in
figure~\ref{lamltot}, by dashed lines.
For the same initial conditions, that is
for $T_0=140$ MeV, we find~\cite{Kuznetsova:2008zr,Kuznetsova:2008hb}
the ratio $\Sigma(1385)/\Lambda_{\rm tot} \approx$
 0.45 at $T \approx 100$ MeV  (and for the entire range 95--135 MeV,
in good agreement with experimental
data~\cite{Adams:2006yu,Salur:2006jq}).

\section{Conclusions}
We find that the resonant hadron states, considering their very  large decay and
reaction  rates, can often interact beyond the chemical and thermal
freeze-out of stable particles. Thus, the observed yield of
resonances is fixed by the physical conditions prevailing at a later
breakup of the fireball matter rather than the production of
non-resonantly interacting hadrons.
This study quantifies the expectation that, in a dense hadron  medium,
narrow resonances are `quenched'\cite{Rafelski:2001hp}.

Despite a  scenario dependent resonance formation or suppression,
the stable particle yields used in the study of chemical freeze-out remain 
almost unchanged, since
all resonances ultimately decay into the lowest `stable' hadron.   Therefore, after a
description, e.g., within a statistical hadronization model  of the yields
of stable hadrons, the understanding of resonance yields is a second, and  separate task
which helps to establish the consistency of our physical understanding of the hadron
production process.

Our results show that the observable ratio
$\Lambda(1520)_{\rm ob}/\Lambda_{\rm tot}$ can be suppressed by two effects. First
$\Lambda(1520)$ yield is suppressed   due to excitation of heavy
$\Sigma^*$s in the resonance scattering process. Moreover, the final
$\Lambda(1520)_{\rm ob}$ yield is suppressed, because $\Sigma^*$s,
which decay to $\Lambda(1520)$, are suppressed at the end of the
kinetic phase evolution by their (asymmetric) decays to lower mass
hadrons.

We  resolve resonance puzzle in that we find that
some resonances can  be enhanced and some suppressed.    Specifically $\Sigma(1385)$
is strongly enhanced, since the dense pion gas especially for 
$\gamma_q>1$ pushes the $\Lambda$ into $\Sigma(1385)$. On the other hand,
the narrow $\Lambda(1520)$ is  depleted by pions pushing it over to
high mass resonances, which later can decay without repopulating  $\Lambda(1520)$.
This effect is particularly strong if we observe that there are fewer high energy particles than
Boltzmann distribution predicts in a rapidly expanding and cooling fireball.
\\

{\it Acknowledgments:}
JR thanks the organizers of WPAC meeting in Krakow for their kind invitation to make this
presentation,  and the excellent ambiance. He also
thanks PD Dr. Peter Thirolf and  Prof. D. Habs, Director of  the Cluster of Excellence
in Laser Physics ---  Munich-Center for Advanced Photonics  (MAP) for their hospitality
in Garching where this research was in part carried out.
This research was supported  by the DFG--LMUexcellent  grant, and by a grant
from: the U.S. Department of Energy  DE-FG02-04ER4131.


\end{document}